\documentclass{aa}
\usepackage{epsfig}
\begin{document}
\newcommand{\beq} {\begin{equation}}
\newcommand{\enq} {\end{equation}}
\newcommand{\DIRBE}{$COBE$-DIRBE}
\newcommand {\chisq}{\chi^{2}}
\def\lsim{\, \lower2truept\hbox{${< \atop\hbox{\raise4truept\hbox{$\sim$}}}$}\,}
\def\gsim{\, \lower2truept\hbox{${> \atop\hbox{\raise4truept\hbox{$\sim$}}}$}\,}

\title{Full sky study of diffuse galactic emission \\ at decimeter 
wavelenghts}
\titlerunning{Diffuse Galactic Emission \\ at Decimeter Wavelenghts}

\authorrunning{P.~Platania et al. }

\author{ 
   P.~Platania \inst{1}, C.~Burigana \inst{2}, D.~Maino \inst{1}, 
   E.~Caserini \inst{1}, M.~Bersanelli \inst{1,3}, B.~Cappellini \inst{1}, 
   A.~Mennella \inst{3}
   }

\offprints{Paola.Platania@mi.infn.it}

\institute{
Dipartimento di Fisica,
Universit\`a degli Studi di Milano, via Celoria 16, I-20133 Milano, Italy\\
\and
IASF/CNR, Sezione di Bologna,
              via P.~Gobetti, 101, I-40129 Bologna, Italy\\
\and
IASF/CNR, Sezione di Milano,
              via Bassini 15, I-20133 Milano, Italy\\
}

\date{Sent February 26, 2003}
\date{Received 5 March 2003 / Accepted 24 July 2003}

\abstract{A detailed knowledge of the Galactic radio continuum is of high
interest for studies of the dynamics and structure of the Galaxy as well
as for the problem of foreground removal in Cosmic Microwave Background
measurements. In this work we present a full-sky study of the diffuse
Galactic emission at frequencies of few GHz, where synchrotron radiation
is by far the dominant component. We perform a detailed combined analysis
of the extended surveys at 408, 1420 and 2326~MHz (by Haslam et al. 1982,
Reich 1982, Reich \& Reich, 1986 and Jonas et al. 1998, respectively). 
Using the technique applied by Schlegel et al. (1998) to the IRAS data, we 
produce destriped versions of the three maps. This allows us to construct a 
nearly-full-sky map of the spectral index and of the normalization factor with 
sub-degree angular resolution. The resulting distribution of 
the spectral indices has an average of
$\bar\beta = 2.695$ and dispersion $\sigma_{\beta} = 0.120$. 
This is representative for the
Galactic diffuse synchrotron emission, with only minor effects from
free-free emission and point sources. 


}

\maketitle

%
\section{Introduction}
The detailed study of radio continuum emission allows a direct evaluation
of important parameters describing the dynamics and structure of the
Galaxy, such as magnetic field distribution and electron energy
distribution. Moreover, diffuse Galactic radio emission is one of the 
main sources of unwanted signal in observations of the Cosmic Microwave
Background (CMB) at $\nu \lsim 50$ GHz.  High precision 
imaging of the CMB fluctuations is a
unique cosmological tool to determine fundamental parameters that describe
the structure and evolution of the Universe and directly probes the
early phase of cosmic history.

The new generation of CMB imaging experiments (in particular the 
$WMAP$~\footnote{http://map.gsfc.nasa.gov/} and
$PLANCK$~\footnote{http://astro.estec.esa.nl/Planck/} 
space missions) call for extended, detailed 
information of Galactic foreground emissions. 
While such exquisite precision  measurements provide a
great information on foreground emissions at microwave frequencies,  
a deep knowledge at radio and sub-millimetre frequencies  
of all emissions that are between us and the Last Scattering Surface where CMB 
photons originate significantly contribute to the accurate  
exploitation of microwave data. In particular great attention has to
be devoted to Galactic foreground emissions with their own power 
spectra and frequency dependence.

The three main physical processes that produce Galactic diffuse emissions
are synchrotron, free-free (or thermal bremsstrahlung) and dust emission.
While free-free does not dominate in any observed
frequency band, synchrotron and dust are the major contributors at low and
high frequencies, respectively. Therefore, it is convenient to deal with
these two foreground emissions using template maps obtained from observations where
such emissions overwhelm the others. In this way it is possible to
cross-correlate experimental CMB data with external data sets
(\cite{banday}, Bennett et al. 2003) to clean CMB signal from foreground 
signal as well as performing component separation with the use of priors on 
foreground emissions (\cite{maino}).

As for the dust emission is concerned, a detailed
work has been done by Schlegel, Finkbeiner \& Davis 
(1998, hereafter SFD)
to combined IRAS with \DIRBE\ data in order to obtain
a template for thermal dust emission with IRAS
angular resolution and \DIRBE\ calibration quality.

In the case of synchrotron emission, observations suffer from many
systematic effects which depend on the specific
experimental setup and observing strategies.
Due to the need of covering the whole sky, many
instruments perform observations that will be finally
combined in a unique data set. 
This introduces problems such as non-uniform zero-levels,
resulting in a normalization of the signal when combining different
patches of observation taken by different instruments
located in different places. Furthermore the involved
technology makes use of radiometer
detectors which typically show $1/f$ noise induced by  
gain fluctuations that leads to striping effects
in the maps when combined with the observing
strategy.

In this paper we address the issue of cleaning
low frequency maps from their
striping effect employing a technique
similar to that used by SFD on the IRAS
plates. We apply this to the three largest low frequency maps:
the 408~MHz map by Haslam et al. (1982),
the 1420~MHz map from Reich (1982) and Reich \& Reich (1986),
and the 2326~MHz map by Jonas et al. (1998).

Our goal is to produce reliable templates 
of spectral index and normalization factor of the power 
law description of the Galactic emission at $\approx$~GHz 
frequencies, by removing the main observation artifacts and 
taking into account
the survey calibration and zero level uncertainties.
As the recent $WMAP$ results show, any template of the Galactic 
emission is frequency dependent (\cite{MAP_fore}). A comparison between the 
synchrotron spectral index
map at $WMAP$ frequencies and that derived from the analysis of
radio surveys at $\approx$~GHz frequencies
will add information on the physical processes
of electron energy loss and the properties of Galactic structures.

This paper is organized as follows. 
In Sect.~\ref{maps} we briefly summarize the main properties of the 
considered surveys. The description of the first steps of the 
adopted data analysis method, source identification and destriping, 
is given in Sect.~\ref{code}, while in Sect.~\ref{beta} we discuss 
the minimization 
of the discontinuities of the maps of spectral index and normalization 
factor of the radio emission power law due to 
the survey calibration and zero level errors 
and present our results.
Finally, we discuss the results and draw our main conclusions
in Sect.~\ref{concl}.

%
%
\section{Data description}
\label{maps}

Synchrotron radiation, arising from cosmic ray electrons spiralizing into the 
interstellar magnetic
field, is dominant over bremsstrahlung (free-free) emission up to 10--20 GHz. 
The emission spectrum has a power law shape, whose spectral index 
varies across the sky reflecting variations of the electrons density and 
magnetic field; a realistic model of the synchrotron emission has to take into
account this significant spatial changing, besides the steepening of the
spectrum with increasing frequency, due to the increase of energy losses by
high energy electrons. 
Free-free emission comes from electron-ion scattering with a power law
shape in frequency; since the diffuse emission at 
$\approx$ GHz frequency is dominated by synchrotron radiation, no free-free 
templates are available at this frequencies. 

The only full sky map at decimeter range is the well known 408~MHz map
(\cite{haslam}). Obtained from four different experiments, the overall map 
has an angular resolution of $0.85^{\circ}$ and a 
full beam brightness sensitivity of $2$ K. 
The estimate of calibration and zero level errors quoted by the authors are 
$10\%$ and 3 K respectively. 

\noindent The 1420~MHz map (\cite{1420MHza}, \cite{1420MHzb}), covering 
the sky at $-19^{\circ}<\delta<90^{\circ}$, has a quoted 
full beam brightness sensitivity of 50~mK  ($0.59^{\circ}$ 
angular resolution). The authors estimate $5\%$ of calibration error and $0.5$ 
K error of the zero level. 

\noindent The higher frequency map (2326~MHz, \cite{2326MHz}), covering 
the sky region  
$-83^{\circ}<\delta<13^{\circ}$ for $0h<RA<12h$ and 
$-83^{\circ}<\delta<32^{\circ}$ for $12h<RA<24h$, has an angular 
resolution of $19'$ and a full beam brightness sensitivity of $25$ mK. 
Calibration and zero level errors are $5\%$ and $80$ mK, respectively. 
The receiver was sensitive to linearly polarized radiation.

The combination of gain drifts and observational strategy results in artificial
structures in the three final maps: these appear like stripes particularly
intense in the scan directions. A careful analysis of the systematic effects in
the 408 and 1420~MHz 
maps has been performed by Davies et al. (1996). Following their considerations,
the most evident artifacts in the 408~MHz map are the nearly vertical stripes at 
constant right ascension generated by the elevation scanning strategy and not 
completely eliminated by the nodding scan technique used in the experiment.  
Horizontal stripes are also evident at declination $<20^{\circ}$.
In the case of the 1420~MHz map the inclined striations reflect the 
azimuthal scanning strategy. Further artifacts show up in the horizontal 
direction mostly in the declination range between $20^{\circ}$ to $50^{\circ}$. 
After point source extraction, the authors remove the major Fourier 
components which produce horizontal and vertical striations on angular scales 
$<5^{\circ}$ in the Haslam map and diagonal and horizontal striations on scales 
$<5^{\circ}$ and $<9^{\circ}$ respectively in the 1420~MHz map. 
As they point out, this technique removes the easily visible angular 
frequency leaving spurious structures on scales of several tens of degrees.
Vertical striations parallel to the scanning direction are visible in the 
2326~MHz map too and remain even after the smoothing applied by Jonas et al. 
(1998). 
%
%
\section{Description of the destriping method}
\label{code}
\begin{figure*}
   \centering
\caption{The block-diagram summarizes the steps of the destriping procedure.}%
\label{block}
\end{figure*}
\begin{figure*}
   \centering
\caption{A patch of the 408~MHz map is shown (left) together with the same 
destriped (right). The lower is the stripe image; note the diagonal features 
together with the dominant vertical striations. For a better qualitative 
comprehension, different color tables have been used for the two upper images 
and the lower one. For the same reason a different color table has been used for 
the visualization of the entire map (Fig.~\ref{fig:Haslam}).}
              \label{fig:patch_15}%
\end{figure*}
To clean the data from the artifacts mentioned above, it is necessary to select the contaminated 
directions and remove the excess signal; for this purpose we adapted the 
method proposed by SFD to destripe the IRAS map. 
We remember that, although a deep comprehension of the 
experimental/observational origin of stripes in the considered 
survey certainly helps the data exploitation, this and 
other kinds of destriping algorithms (see e.g.  
Keih\"anen al. 2003, and references therein, 
for a different destriping approach) 
have the remarkable property to work almost independently 
from the details of the stripe pattern.
For a comprehensive explanation of the algorithm concept we refer to their 
original paper; here we explain briefly the main steps of the process, that 
are also summarized in the block diagram of Fig.~\ref{block}. 
\noindent The analysis is not performed on the map as a whole, since the code
works on ``squared'' patches. We divided the maps of the surveys at 408, 1420, 
and 2326~MHz respectively
into 44, 33, and 22 partially overlapping patches of $152 \times 152$ pixels 
with angular size $\Delta \theta \times \Delta \phi = 0.35^{\circ} \times 
0.35^{\circ}$.
The destriping code acts on maps gridded in standard polar coordinates
$\theta , \phi$ by using an Equi-Cylindrical Projection 
(ECP) pixelization, i.e. a rectangular, not equal area, projection of the 
sphere. 

\noindent The algorithm first performs a point source removal, selecting regions 
whose signal exceeds the median filtered image by a given threshold.
The source signal is then 
replaced with the median signal from an anulus around the source
(SFD);
this allows destriping to be performed in the Fourier
space, where power from bright sources may add and overwhelm many wavenumbers.

\noindent The Fourier domain, parameterized in polar coordinates 
($k_{\rm{r}}$,$k_{\theta}$),
is then divided into 90 bins of $2^{\circ}$ in $k_{\theta}$ and 
a background power $P_{\rm{back}}$ is calculated by ``finding the median at each 
$\theta$ and then median filtering the result'' (SFD). 
The contaminated wavenumbers are those for 
which the power ratio $\gamma_{\theta}$ between the power, $P_{\theta}$, in 
each $k_{\theta}$
bin and the background power, $P_{\rm{back}}$, differs significantly from unity.     
The difference, $\Delta P_{\theta}$, between the power as a function 
of $\theta$ and a mean ``good'' power, $P_{\rm{ave}}$ 
(obtained by averaging all power 
at each ($\kappa_{\rm{r}}$, $\kappa_{\theta})$ with $\gamma_{\theta}$ lower than the 
stripes threshold  $\bar{\gamma}$), 
is then inversely transformed to get the image of stripes only, to be 
subtracted from the original patch.

\noindent
The original resolution of the 2.326 GHz map is approximately the same 
as the pixel size used in this analysis; we tested the influence of a better 
sampling on the destriping procedure adapting the algorithm to the same map 
with $2048 \times 1024$ pixels (each pixel of $0.175^{\circ} \times 
0.175^{\circ}$). The results show that the pixel by pixel temperature 
difference between the same destriped patches with different pixel size is, on 
average, less than 5\% for 70-80\% of the pixels. Given that 
the three frequencies maps will be jointly used in the following analysis, we 
choosed to perform the destriping on the 2.326 GHz map with the same pixel size
as the others.

\noindent The real space image of a typical Haslam patch is shown in 
Fig.~\ref{fig:patch_15} together with the same patch after destriping. 
In the same figure the real space image of the subtracted stripes 
is shown.
The destriped patches are then joined together to obtain the destriped 
map\footnote{For three patches of the Haslam map the procedure failed in
finding a lower signal anulus to replace the source signal; in these cases the
algorithm did not perform destriping.}. 
\subsection{Determination of the thresholds}
\label{thresh}

\begin{figure*}
   \centering
	\caption{Ratio between the power, $P_{\theta}$, in each $k_{\theta}$
bin and the background power, $P_{\rm{back}}$, for the same patch of 
Fig.~\ref{fig:patch_15} as a function of the angle $\theta$. 
Different thresholds $\bar\gamma$ are shown (horizontal dashed lines). 
The highest value selects only very intense stripes; by lowering the 
threshold, lower peaks are also included.}
              \label{fig:gamma_theta}%
\end{figure*}

\begin{table*}
\begin{center}
\footnotesize
\begin{tabular}{|c| c c c c|}
\hline
~          & $\delta T <1\%$  & $1\%< \delta T <5\%$ & $5\%< \delta T <10\%$ & $\delta T >10\%$ \\
\hline
408      	& 36.4$\%$	   &  	55$\%$	 	 & 	7.0$\%$ 	  & 0.6$\%$ 			\\
1420 		& 89.14$\%$ 	   &  	10.7$\%$	 & 	0.15$\%$	  & 0.01$\%$  			\\
2326 		& 18.3$\%$ 	   & 	50.6$\%$	 & 	22.5$\%$	  & 8.6$\%$ 			\\
\hline
\end{tabular}
\end{center}
\caption{Distribution of percentage temperature variation $\delta T$ values in 
four different ranges and the corresponding pixel percentage. $\delta T$ is 
evaluated as the difference after and before destriping.}
\label{tab:stripes}
\end{table*}

Criteria for the choice of the source extraction threshold $h_{\rm{s}}$ and the stripes
threshold $\bar{\gamma}$ have to be carefully addressed in order to optimize 
the destriping procedure. 
We performed several tests to determine the best thresholds for source and 
stripe removal and to estimate the relative errors and the effect on the 
destriped
maps. 
The choice of the thresholds does not have the same effect on 
every patch of one map. However, the values of $h_{\rm{s}}$ and $\bar{\gamma}$ have 
been mantained fixed for all the patches of the same map, since the division of
 each map in patches is useful in the analysis but does not reflect a real 
experimental situation. We therefore look for the values that give the most
uniform results on the whole map.

A first test for the effectiveness of the source removal procedure is 
to look at the cleaned image to check for the presence of residual sources. 
It should be noted that the optimal choice for stripe threshold $\bar{\gamma}$ 
is 
strictly related to source removal. Reminding that the goal is to obtain an 
image of stripes to be subtracted to the original patch, two opposite 
non-optimal 
choices of the source threshold $h_{\rm{s}}$ could affect the destriping
method. The choice of a high threshold $h_{\rm{s}}$ (at limit, avoid source subtraction
 at all)
implies that power from unsubtracted sources could be attributed to stripes and 
then uncorrectly removed from the original map.  
On the other hand, the choice of a too low threshold $h_{\rm{s}}$ implies
that not only the signal from point sources is subtracted but also power from 
stripes that will not be removed in the destriping procedure.

To optimize the stripe removal, in the Fourier space we first 
eliminate the 
peaks higher than $\bar{\gamma}$ as a function of the direction 
$\kappa_{\theta}$ (see Fig.~\ref{fig:gamma_theta}). 
Stripes have a high level of directionality and this gives a good criterion 
for the choice of the threshold value: $\bar{\gamma}$ has to be not too 
low in order to maintain this feature and avoid subtraction of real astrophysical 
signals. On the other side, a too high $\bar{\gamma}$ would eliminate 
only stripes with very high signal, often in one single direction. 
Fig.~\ref{fig:gamma_theta} shows the pattern of the ratio 
$P_{\theta}/P_{\rm{back}}$ as a function of the angle $\theta$
for the same patch of Fig.~\ref{fig:patch_15}.

After the identification of the optimal stripe threshold $\bar{\gamma}$, we 
tested the influence of this choice on the destriped map, comparing it with 
the same map destriped with a higher value of $\bar{\gamma}$. By changing 
$\bar{\gamma}$ by $20\%$ (i.e. to a value well beyond a reasonable choice of 
the threshold), the number of pixels for 
which the temperature changes more than $10\%$ are less than $\simeq 1\%$ in 
the 408~MHz map, $\simeq 0.1\%$ at 1420~MHz, and reaches a few 
percent in the 2326~MHz map (each fraction is referred to the sky coverage of 
each survey).
As shown in Table~\ref{tab:stripes} the destriping procedure does not 
change the intensity level of the surveys except for a small percentage of
the pixels for the 408 and 1420~MHz maps and for less than 10\% of the pixels 
for the 2326~MHz map.
%
\subsection{Resulting maps}
\label{res_maps}
\begin{figure*}
   \centering
\caption{Top: original 408~MHz map (\cite{haslam}). Bottom: destriped 408~MHz
map (this work).}
              \label{fig:Haslam}%
\end{figure*}
\begin{figure*}
   \centering
	\caption{Top: original 1420~MHz map
	(\cite{1420MHza}, \cite{1420MHzb}). 
	Bottom: destriped 1420~MHz map 
	(this work).}
              \label{fig:Reich}%
\end{figure*}
\begin{figure*}
   \centering
	\caption{Top: original 2326~MHz map (\cite{2326MHz}). 
	Bottom: destriped 2326~MHz map (this work). 
For an easier division in relatively 
large squared patches, the region $13^{\circ}<\delta<32^{\circ}$
of the 2326~MHz map has not been included in the data analysis steps 
described in Sects.~3.1 and 3.2.}
              \label{fig:Rhodes}%
\end{figure*}
The resulting maps are shown in Figs.~\ref{fig:Haslam}, \ref{fig:Reich}, 
\ref{fig:Rhodes}. 
We also produced the three destriped maps without point sources using
the same technique that has been implemented in the destriping algorithm. The 
technique works well for isolated sources; in the case of extended structures, 
however, the replacement of the high source signal with the median background 
cause the smoothing of the structure borders. We thus preferred to 
leave the sources in the maps; their contribution to the spectral index 
distribution will be discussed in Sect.~\ref{results}.
\noindent Maps are also available in HEALPix\footnote{\tt
http://www.eso.org/science/healpix} format (G\`orski et al. 1999)
with $N_{\rm side}=256$.
The conversion from ECP to HEALPix maps is performed working in real
space via an intermediate step, i.e. an HEALPix map with higher
resolution ($N_{\rm side}=4096$, the maximum available within HEALPix
packages).
For each HEALPix pixel of the
higher resolution map, the conversion code\footnote{Details on the conversion 
procedure can be found in Cappellini et al. (2003a, 2003b).} 
finds the corresponding ECP pixel and assign its signal value.
The map is finally degraded to $N_{\rm side}=256$.
%
%
\section{Spectral index and normalization factor maps}
\label{beta}
\begin{table*}
\begin{center}
\footnotesize
\begin{tabular}{|c| c c c c|}
\hline
Survey          & calibration      & Zero level   & Baseline & Extragalactic 	\\
Frequency (MHz) & error (per cent) & error (K)   & correction & background (K) (K) 	\\
\hline
408      	& 10 		   & 3 		 & -	  &  5.92			\\
1420 		& 5 		   & 0.5 	 & $-$0.13	  & 2.83		\\
2326 		& 5 		   & 0.080 	 & -	  &  	2.75		\\
\hline
\end{tabular}
\end{center}
\caption{Calibration errors $\sigma_{\rm cal}$, zero level errors $\sigma_{\rm 0l}$ and 
baseline corrections from the original experiments at 408, 142 and 
2326~MHz; 
extragalactic background for the 408 and 1420~MHz maps have been estimated by 
Lawson et al. (1987), for the 2326~MHz map in this work following the
same procedure.}
\label{tab:surveysdata}
\end{table*}

\noindent One of the main goals of this work is the study of the spectral
index of the Galactic diffuse emission over the whole sky, except the region
$-90^{\circ}<\delta<-83^{\circ}$ where only 408~MHz data are available. 
Before evaluating the distribution of the spectral index $\beta$, 
we subtracted the unresolved extragalactic background and CMB absolute 
temperature at 408 and 1420~MHz following Reich \& Reich (1988). 
The unresolved extragalactic background for the 2326~MHz map has been estimated 
using the method proposed by Lawson et al 1987. In addition a baseline offset of
$-0.13$~K for the 1420~MHz map has been considered (see Table 
\ref{tab:surveysdata}). 
The two maps at higher frequencies have been convolved to match the Haslam map
resolution. 

We carefully considered the statistical and systematic errors of each surveys 
in the analysis; in the definition of the overall temperature error for each 
pixel, we include 
the quoted sensitivity, $\sigma_{\rm sens}$, the errors on the zero level, 
$\sigma_{\rm 0l}$, and calibration, $\sigma_{\rm cal}$, and a correction factor, 
$c_{\rm a}$, to account for the different area of the pixels
\begin{equation}
c_{\rm a} = L \sqrt{\sin \theta_i}, 
\end{equation}
where $L$ is the pixel side in steradian and $\theta_i=\pi - L j$ is the 
colatitude of the position of $j-th$ pixel lower border. 
The convolution of the 1420 and 2326~MHz maps to the resolution of 
the 408~MHz map has an impact on the sensitivity; we took it into account in
the error budget.
The experiment that produced the 2326~MHz map was sensitive to a linear 
polarization direction (\cite{2326MHz}), while the other two surveys 
give the total (non polarized) intensity.
The available polarization surveys at low frequency do not cover large regions
of those observed by the Rhodes/HartRAO 2326~MHz telescope
(for example \cite{d97} and 1999) or are undersampled (\cite{bs}); thus a 
pixel by pixel subtraction of the linear polarized component of the 2326~MHz 
map based on the polarization data is not feasible. We prefer to account for 
this discrepancy between the maps including the conservative term 
$\sigma_{\rm pol, 2326}=20\%$ in the
error evaluation of the 2326~MHz temperature (\cite{bacci}). 
The final expressions for the overall errors at the three frequencies are:
%


\begin{eqnarray}
   dT_{\rm 408} & = & \Big[(\sigma_{\rm sens, 408}\theta^b_{\rm 408}/c_{\rm a})^2+ \nonumber \\
   	    & ~ & \delta_{\rm 0l,408}^2+(T_{\rm 408} g_{\rm cal,408})^2\Big]^{1/2} \nonumber \\
   dT_{\rm 1420}& = & \Big[(\sigma_{\rm sens,1420}\theta^b_{\rm
   1420}(\theta^b_{\rm 1420}/\theta^b_{\rm 408})/c_{\rm a})^2+ \nonumber \\ 
            & ~ & \delta_{\rm 0l,1420}^2+(T_{\rm 1420} g_{\rm cal,1420})^2 \Big]^{1/2} \nonumber\\ 
   dT_{\rm 2326}& = & \Big[(\sigma_{\rm sens,2326}\theta^b_{\rm
   2326}(\theta^b_{\rm 2326}/\theta^b_{\rm 408})/c_{\rm a})^2+ \nonumber \\
            & ~ & \delta_{\rm 0l,2326}^2+(T_{\rm 2326}g_{\rm cal,2326})^2 +  \nonumber\\
            & ~ & (T_{\rm 2326} \sigma_{\rm pol,2326})^2 \Big]^{1/2} 
\label{eq:T_errors}
\end{eqnarray}
where $\theta^b$ is the full width half maximum of the experiment at each 
frequency.
Reminding that the relation between synchrotron radiation temperature and 
frequency is a
power law,  
\begin{equation}
	T_{\nu}=a {\nu}^{-\beta} \,,
\end{equation}
and
\begin{equation}
	\mathrm{log} T_{\nu}=\alpha -\beta \rm{log}  \nu \,,
\end{equation}
where $\mathrm{log} a = \alpha$, the spectral index 
and the normalization factor and their errors are simply given by: 
\begin{equation}
	\beta_{\nu_{\rm 1}/\nu_{\rm 2}} = -\frac{\mathrm{log}(T_{\nu_{\rm
	1}}/T_{\nu_{\rm 2}})}
	{\mathrm{log}(\nu_{\rm 1}/\nu_{\rm 2})} \,,
\end{equation}
\begin{equation}
\alpha= \frac{\mathrm{log}\nu_{\rm 1} \mathrm{log}T_{\nu_{\rm
2}}-\mathrm{log}\nu_{\rm 2} \mathrm{log} T_{\nu_{\rm 1}}}
{\mathrm{log} \nu_{\rm 1}-\mathrm{log} \nu_{\rm 2}} \, ,
\end{equation}
\begin{equation}
\label{beta_error}
\sigma_{\beta} = |\mathrm{log}   \nu_{\rm 1}-\mathrm {log}\nu_{\rm 2}|^{-1}
               \sqrt{\Big(\frac{dT_{\rm 1}}{T_{\rm 1} \mathrm {log}10}\Big)^2+
	       \Big(\frac{dT_{\rm 2}}{T_{\rm 2} \mathrm {log}10}\Big)^2}
\end{equation}
\begin{eqnarray}
\sigma_{\alpha} & = & |\mathrm {log}\nu_{\rm 1}-\mathrm {log}\nu_{\rm 2}|^{-1} \nonumber \\  
                   & ~ & \times \sqrt{{\nu_{\rm 2}}^2 \Big(\frac{dT_{\rm
		   1}}{T_{\rm 1} \mathrm{log}10}\Big)^2+{\nu_{\rm 1}}^2 
                         \Big(\frac{dT_{\rm 2}}{T_{\rm 2} \mathrm {log10}}\Big)^2} \, .
\end{eqnarray} 
where 1 and 2 are two generic frequencies.

In the central region where the three maps overlap, we perform a least square 
fit to evaluate the $\beta$, $\alpha$ and their errors (\cite{num_rec}) 
assuming a constant $\beta$ between 408 and 2326~MHz.
We repeated the same analysis including only the statistical errors in the 
error budget. 
\noindent We estimated the influence of the stripes threshold choice on the 
spectral index 
distribution: the mean error is $\sim 1\%$, thus negligible in the error budget.  

As we expect, the average spectral index and 
dispersion are not significantly affected by destriping, while substantial 
changes are produced on a pixel by pixel basis (up to $\Delta \beta \sim 0.4$). 
Fig.~\ref{fig:stripes+beta} shows the difference $\Delta \beta$ between the 
spectral indices evaluated with destriped and original maps.

\begin{figure*}
   \centering
\caption{Map of the differences $D\beta$ between the spectral index evaluated 
with the destriped and the original maps. Features corresponding to stripes
are evident.}
              \label{fig:stripes+beta}%
\end{figure*}

%
\subsection{Frequency variation analysis} 
\label{subsect:freq_var}

\begin{figure*}
   \centering
	\caption{Distribution of the differences 
	$\beta_{\rm 408,1420}-\beta_{\rm 1420,2326}$. The binsize is 0.01.}
              \label{fig:hist_diff}%
\end{figure*}

The steepening of the synchrotron spectrum with increasing frequency 
has to be taken into account in any estimate of the Galactic radio emission. 
We tested the assumption that the spectral index between 408 and 2326 MHz does 
not change significantly when evaluated with different combinations of the 
three surveys. 
In the overlapping region, where data at the three frequencies are available, 
we compare $\beta_{\rm 408,1420}$ with $\beta_{\rm 1420,2326}$.
Fig.~\ref{fig:hist_diff} shows the distribution of the difference between the 
spectral index evaluated with 408 and 1420 MHz data, 
$\beta_{\rm 408,1420}$, and the 
spectral index evaluated with the 1420 and 2326 MHz data, $\beta_{\rm 1420,2326}$. 
The shape of the distribution indicates that there is not a 
clear steepening of $\beta$ when using these surveys. Moreover, we compare 
pixel by pixel the distribution of the difference with the standard deviation 
$\sigma_{\beta}$, where, for
each pixel $\sigma_{\beta}^2=\sigma_{\rm \beta(408,1420)}^2+
\sigma_{\rm \beta(1420,2326)}^2$: only for the $5\%$ of the pixels the difference 
between the spectral indices is larger
than the estimated error. The same analysis performed with the statistical
errors only shows that for the $25\%$ of the pixels the difference is larger
than $\sigma_{\rm \beta, stat}$.  
We conclude that, generally, the estimated error is larger than the 
frequency variation of the spectral index $\beta$ between 1420 and 2326~MHz.
This motivates the idea to jointly use, where possible, all the three
surveys to derive a single map of spectral indices and normalization
factors in the frequency range 0.408--2326~MHz.
As discussed in Sect.~\ref{maps}, these surveys are affected by 
zero level and calibration uncertainties; the latter are 
particularly relevant at 2326~MHz. 
In the next subsection we discuss an attempt to reduce, at least
in part, the impact of these uncertainties in the joint analysis
of the three surveys.
 


%
\subsection{Discontinuities} 
\label{subsect:disc}
\begin{figure*}
   \centering
	\caption{Differences in $\beta$ in the case of statistical error only
	(a) and systematics included (b); pixel number [0-1024] corresponds to
	right ascension [$-180^{\circ}$, $180^{\circ}$]. The curves show the up (a1 and b1) 
	and down (a2 and b2) discontinuities before (green) and after (purple) 
	minimization. It is worth to remind that the values $\Delta\beta$ and 
	$\Delta\alpha$ before minimization depend on the kind of errors 
	included in the analysis, via their influence on the fit in the case 
	we use data at all the three frequencies.}
              \label{fig:disc_beta}%
\end{figure*}
\begin{figure*}
   \centering   
	\caption{Differences in $\alpha$ in the case of statistical error only
	(a) and systematics included (b). The curves show the up (a1 and b1) 
	and down (a2 and b2) discontinuities before (green) and after (purple) 
	minimization. See comments for Fig. ~\ref{fig:disc_beta}.}
              \label{fig:disc_alfa}%
\end{figure*}

\noindent As one might expect, the $\beta$ and $\alpha$ maps show two horizontal
lines of discontinuity corresponding to
the borders of the maps at 1420 and 2326~MHz. 
Of course these discontinuities arise from measurements and 
inter-calibration errors and do not have any astrophysical meaning.

\noindent A smoothing of the data to a lower
resolution can be easily applied to eliminate the appearence of this effect but
would not eliminate its causes.
The same effect has been observed by Giardino et al. (2002) who 
combined the 408, 1420 and 
2326~MHz maps to obtain a large angular resolution ($10^\circ$) map of $\beta$.
To eliminate the discontinuities, they tried different techniques looking for a
baseline offset for one data set, with no improvement. 
In the overlapping region, they set the spectral index to a 
weighted average of $\beta_{\rm 408,1420}$ and $\beta_{\rm 408,2326}$; the choice of the 
weights ensures that the three different sky regions smoothly merge into each 
other. By convolving the data sets to $10^\circ$ resolution, the stripes and 
discontinuities effects were averaged out.

\noindent 
In this work we are interested in the high resolution information of the 
spectral index; after the destriping procedure discussed in Sect.~\ref{code},
we investigate to what extent the discontinuities in the $\alpha$
and $\beta$ maps could be accounted for within the quoted measurements 
systematic errors of the three surveys (see Table~\ref{tab:surveysdata}), 
i.e., gain and zero-level errors.  
One should be aware that these uncertainties include a number of effects
related to the details of the scanning strategy and calibration procedure
adopted in the observations. All three surveys are obtained by merging together
different sky patches with zero levels somewhat different from one another.
In the merging phase of the map production and as a-posteriori
analysis, attempts were made to minimise discontinuities between different 
patches (e.g., Reich \& Reich 1988). In addition, these surveys have been 
calibrated by convolution to sky horn absolute measurements, typically obtained 
at fixed declinations. With increasing distance from the absolute calibration 
data, the uncertainty in the zero level could increase due to residual 
baseline gradients. Therefore using zero-level errors as ``baseline errors'' is
equivalent to the assumption that the underlying constant component is of the 
same order as the total uncertainty. Although not ideal, we follow this 
assumption since it is the only practical possibility at hand. 
A deeper analysis would require the availability of the original raw 
data of each survey to perform a complete reanalysis, a task which is outside 
the scope of this work.

We searched for the combination of six 
parameters (three pairs of parameters, calibration factor $g_{\rm cal}$ and zero 
level correction $\delta_{\rm 0l}$, each for
each frequency map) to rescale the maps 
to account for their zero level and calibration errors with the aim to 
minimize the difference across the discontinuities. Of course, the 
assumption that these three couples of parameters are constant over the 
entire survey is an
over-simplification; in fact it is very hard, if not impossible, to 
recover the real 
pattern of the variations of these parameters from the considered data products 
(see below for an attempt in this direction).  
The six parameters influence the 
$\beta$ and $\alpha$ values through the optimized temperatures of the maps:
\begin{equation}
T_{\rm \nu \, opt} = T_{\nu} g_{\rm cal} + \delta_{\rm 0l} \, .
\end{equation}
We adopted the following $\chi^2$-like form of the functional to be minimized: 
\begin{eqnarray}
	\chi^2 = \sum_{i=1}^{N}\frac {(\beta_{\rm 2d}(i)-\beta_{\rm 3d}(i))^2}
		{\sigma^2_{\rm \beta_{2d}}(i)+\sigma^2_{\rm \beta_{3d}}(i)} + 
		 \frac {(\beta_{\rm 2u}(i)-\beta_{\rm 3u}(i))^2}
		{\sigma^2_{\rm \beta_{\rm 2u}}(i)+\sigma^2_{\rm \beta_{\rm 3u}}(i)} +
		\nonumber \\
		+ \frac {(\alpha_{\rm 2d}(i)-\alpha_{\rm 3d}(i))^2}
		{\sigma^2_{\rm \alpha_{\rm 2d}}(i)+\sigma^2_{\rm \alpha_{\rm 3d}}(i)} + 	
		\frac {(\rm \alpha_{\rm 2u}(i)-\alpha_{\rm 3u}(i))^2}
		{\sigma^2_{\rm \alpha_{\rm 2u}}(i)+\sigma^2_{\rm \alpha_{\rm 3u}}(i)} \, ,  	
\label{eq:chisqr}
\end{eqnarray}
where $u$ and $d$ refers to the $\it{up}$ and $\it{down}$ discontinuities respectively,
the low indices 2 and 3 refers to $\beta$ and $\alpha$ evaluated by exploiting 
two or three frequency maps, and the sum is carried out over 
$i-th$ azimuthal position of all the N pixels 
of the top and bottom border lines of the overlapping region.
The minimization has been carried out by using the MINUIT package of 
the CERN library~\footnote{http://cern.web.cern.ch/CERN/}.
\noindent Different approaches have been tested in this minimization: 
we left all the parameters free, we kept fixed each of them 
(e.g. $\delta_{\rm 0l}=0$ and
$g_{\rm cal}=1$ or viceversa for a given frequency map) 
while leaving the others free; we varied each of them at 
one time keeping fixed the others, and finally, we varied the 
parameters $\delta_{\rm 0l}$ and $g_{\rm cal}$ of only a given frequency map by 
keeping the ``original'' parameters at the other 
frequencies~\footnote{Of course the choice $g_{\rm cal}=1$ and $\delta_{\rm 0l}=0$ corresponds
to the ``original'' map.}.
\begin{table}
\begin{center}
\footnotesize
\begin{tabular}{|l| c c |}
\hline
~                 		& statistical 	& statistical + systematic   	\\
~		  		& errors only	& errors 		\\
\hline
$g_{\rm cal,408}$  		& 0.877 	&  0.973		\\
$g_{\rm cal,1420}$ 		& 1.100 	&  1.066		\\
$g_{\rm cal,2326}$ 		& 0.827 	&  0.793		\\
$\delta_{\rm 0l,408}$	& 3.432~K   	&  0.676~K   		\\
$\delta_{\rm 0l,1420}$	& -0.0410~K   	& -0.0267~K   		\\
$\delta_{\rm 0l,2326}$	& 0.0366~K   	&  0.0357~K   		\\
$\chi^2$/d.o.f.		& 14.381	& 0.204		\\
\hline
\end{tabular}
\end{center}
\caption{Calibration factors and zero level corrections for the three maps as 
recovered after the minimization for statistical errors only and with 
systematic errors includes. The $\chi^2$/d.o.f. is also shown.}
\label{tab:parameters}
\end{table}
By observing that two frequency maps (at 408 and 2326~MHz) show, before 
destriping, 
almost vertical striations due to the scanning direction, we implemented also a
 modified 
version of this minimization technique: instead  of searching for six overall 
parameters
over the whole maps, we searched for six azimuthal 
(e.g. as function of the right ascension) profiles of the three
pairs of $\delta_{\rm 0l}$ and $g_{\rm cal}$ by implementing, 
at each $i-th$ azimuthal position, an overall minimization of
the discrepances between the values obtained for $\beta$ and $\alpha$
by using three frequency maps or two combinations of only two frequency maps
for all the pixels at the same $i-th$ azimuthal position but
at the different declinations 
where three frequency maps
are available ($\simeq 90$ pixels at each $i-th$ azimuthal position). 
This approach should in principle better account for possible
azimuthal variations $\delta_{\rm 0l}$ and $g_{\rm cal}$ over the maps
in the case in which they do not show a significant declination dependence.
Unfortunately, this method, although quite effective in reducing 
horizontal discontinuities, introduces undesired vertical 
discontinuities, suggesting for complex patterns of $\delta_{\rm 0l}$ 
and $g_{\rm cal}$ whose recovery cannot clearly be derived by using
such kinds of minimization approaches.

We then adopted the overall minimization approach described by  
the eq.~(\ref{eq:chisqr}) searching for the optimized set of the above six 
parameters. 

Of course, the best result from the point of view of the reduced $\chi^2$ 
and of the minimization of discontinuities, 
is obtained leaving the all six parameters free to vary. 
On the other hand, it is clear that we need to keep them inside their 
corresponding experimental quoted errors. 
This requirement can be relaxed for the parameter $g_{\rm cal}$ referring to 
the map 
at 2326~MHz because of the uncertainty related to the polarization, as discussed
in the previous subsection, which could account for variations of 
$\approx 10$\%.
Moreover, the 2326 MHz data have been calibrated using 2 GHz data 
(Bersanelli et al. 1994) scaled with a spectral index of 2.75; the 
relaxed requirement can take into account also the bias introduced in the 
determination of the spectral index distribution by the calibration procedure.  

We carried out the above minimization with 
$\delta_{\rm 0l}$ and $g_{\rm cal}$ inside their quoted $2 \sigma$ errors
for the frequency maps at 408 and 1420~MHz, and
with $\delta_{\rm 0l}$ ($g_{\rm cal}$) inside its quoted $2 \sigma$ error
($5 \sigma$ error) for the frequency map at 2326~MHz.

We have repeated this analysis by including or not
the contribution of systematics in the error budget,
as discussed in the previous subsection.

In addition, we have tested that the results do not depend on the adopted 
range of allowed parameter values when the systematic errors 
are included. 
On the other hand, neglecting the systematic errors, 
the minimization tends to give a best fit value of 
$g_{\rm cal}$ at 1420~MHz out of its quoted $2 \sigma$ error. 
In this case, we have then repeated the minimization
by constraining $g_{\rm cal}$ at 1420~MHz within different 
ranges (within 1, 2 and 5~$\sigma$) and 
keeping the allowed ranges for the other variables as before.
We find an impact of a few percent on average and always less
than 10\% (with a change less than 1\% for half of the pixels)
on the maps of $\beta$ and $\alpha$. 
In Table~\ref{tab:parameters} we show the six parameters after 
minimization.
The results obtained by including or not systematic 
errors are in agreement within $1 \sigma$ error in each variable of the fit.
The parabolic errors quoted for each recovered parameter 
from the minimization code are much smaller than the corresponding experimental 
uncertainties and discrepancies found including or not systematics, 
and we avoid to report them, being clearly 
``unphysical''~\footnote{Clearly, in the maps of 
the final error on $\beta$ and $\alpha$ the contribution 
of these parabolic errors is also negligible.}.
As expected, the $\chi^2$/d.o.f. is larger (smaller) than unity
when systematic errors are neglected (included) in the analysis.
It is worth to point out that the parameters in 
Table~\ref{tab:parameters} have to be considered not as the 
``real'' correction to the experimental values 
of baseline and calibration of each survey, individually considered, 
but as the result of a procedure devoted 
to obtain more physically meaningful distributions  
of spectral index and normalization 
factor from the joint exploitation of the three surveys. This is particulary 
clear in the case of the 2326~MHz map where, to allow the comparison of a 
linearly polarized survey with the intensity data, the recovered calibration
parameter has been found well out the experimental values.

The differences, $\Delta\beta$ and $\Delta\alpha$, 
in correspondence to the discontinuities,
before and after having applied the described minimization
neglecting or including the systematics, 
are shown in Figs.~\ref{fig:disc_beta} and \ref{fig:disc_alfa}. 
Note that the differences are evaluated between the two adjacent lines across 
each discontinuities, while the minimization is carried out with the values of 
$\beta$ evaluated with two or three frequencies in each border line of the 
overlapping region.
The algorithm has the same qualitative effect if we consider only 
statistical or also systematic errors. 
\subsection{Results}
\label{results}
The resulting maps of $\beta$, $\alpha$ and their errors are shown in 
Figs.~\ref{fig:beta_corr}--\ref{fig:dalfa_corr}. 

From Figs.~\ref{fig:disc_beta} and \ref{fig:disc_alfa} it is
clear that discontinuities do not completely disappear; they are 
reduced, but the minimization
procedure is not able to account for the full complexity of the real 
experimental drifts of the six parameters. 
\noindent The increasing of $\sigma_\beta$ and $\sigma_\alpha$ with 
declination are explained
considering the correction factor $c_{\rm a}$ appearing in 
Eq.~(\ref{eq:T_errors}) to account for non-equal area pixels. 

The distributions of the spectral index $\beta$ are shown in 
Figs.~\ref{fig:isto_onlystaterr} and ~\ref{fig:isto_allerr}; we evaluated the average spectral
index and the dispersion before and after minimization, as summarized in 
Table~\ref{tab:fit_beta};
the effect of minimization is to slightly change both parameters. 
The values well agree with those found by Giardino et al. (2002) of 
$\bar\beta_{\rm 480,1420} = 2.78 \pm 0.17$ and $\bar\beta_{\rm 480,2326} = 2.75 \pm 0.12$ .
The minimization procedure does not alter significantly the pixel by pixel 
values of $\beta$: typically, a variation greater than $0.1$ occurs for a few 
$\%$ of the pixels.

We have estimated the contribution of point sources on the $\beta$ distribution
as follows: we subtracted the point sources from the destriped maps (with the
same technique explained in Sect.~\ref{code}) and performed all the same 
steps on
the analysis. The main results is a sensible reduction of the low spectral 
index tail in the distribution\footnote{The high $\beta$ tail comes from extended sources not removed by the source
selection method.}; on the other hand, when extended regions are
considered, this procedure can sensibly smoothing the outer bordes of the
structure. 
 
The incidence of free-free diffuse contribution has been estimated as follows.
We scaled the $WMAP$ free-free template (Bennett et al. 2003) at 408, 1420 
and 2326~MHz and subtracted from the destriped maps. We obtained a variation of
$\beta$ larger than $0.1$ for less than $8\%$ of the pixels; this difference is 
inside the deviations illustrated in Table~\ref{tab:fit_beta} except for a few 
percentage of the pixels. 
Thus, we can conclude that the diffuse free-free emission does not sensibly
affect the distributions obtained in this work.

We can say that the diffuse Galactic maps resulting from this work are
representative for synchrotron spectral index distribution and 
synchrotron normalization factor.
Point sources contribution is limited to a small number of pixels that could 
be isolated and removed to extrapolate the maps at high frequency and obtain 
synchrotron emission templates.

\begin{table*}
\begin{center}
\footnotesize
\begin{tabular}{|l| c| c c|}
\hline
~                 &	& statistical 	& statistical + systematic   	\\
~		  &	& errors only	& errors \\
\hline
before minimization  & $\bar\beta \pm \sigma_{\beta}$ & $2.742 \pm 0.145$ & $2.734 \pm 0.124$ \\
~		     & \% pixels with $\beta < 2$     & 	0.031	  &  0.015  \\ 
~ 		     & \% pixels with $\beta > 3.5$   & 	0.066	  &  0.05  \\   
\hline
after minimization   & $\bar\beta \pm \sigma_{\beta}$ &	$2.744 \pm 0.122$ &  $2.695 \pm 0.120$	\\
~		     & \% pixels with $\beta < 2$     & 	0.022	  &  0.009   \\ 
~		     & \% pixels with $\beta > 3.5$   & 	0.012	  &  0.012   \\
\hline
\end{tabular}
\end{center}
\caption{Mean values and dispersion of the spectral index distributions. The
percentage of pixels with $\beta <2$ and $>3.5$ is shown.}
\label{tab:fit_beta}
\end{table*}
\begin{figure*}
   \centering   
	\caption{Maps of the synchrotron spectral index $\beta$ after 
	minimization. Top:
the analysis has been carried out with temperature statistical errors only.  
Bottom: systematic errors have been included.}
              \label{fig:beta_corr}%
\end{figure*}
\begin{figure*}
   \centering   
	\caption{Maps of the normalization factor $\alpha$ after minimization. 
	Top:
the analysis has been carried out with temperature statistical errors only.  
Bottom: systematic errors have been included.}
              \label{fig:alfa_corr}%
\end{figure*}
\begin{figure*}
   \centering
	\caption{Maps of the synchrotron spectral index errors $\sigma_{\beta}$ after 
minimization. Top: the analysis has been carried out with temperature 
statistical errors only. Bottom: systematic errors have been included.}
              \label{fig:dbeta_corr}%
\end{figure*}
\begin{figure*}
   \centering
	\caption{Maps of the normalization factor errors $\sigma_{\alpha}$ 
	after 
minimization. Top: the analysis has been carried out with temperature 
statistical errors only.  Bottom: systematic errors have been included.}
              \label{fig:dalfa_corr}%
\end{figure*}
\begin{figure*}
   \centering
	\caption{Analysis with statistical errors only. The histogram of the 
	spectral indices before (a) and after (b) minimization are 
        shown together 
	with their average and dispersion.}
              \label{fig:isto_onlystaterr}%
\end{figure*}
\begin{figure*}
   \centering
	\caption{Analysis with statistical and systematic errors. The histogram 
	of the spectral indices before (a) and after (b) minimization  
        are shown together 
	with their average and dispersion.}
              \label{fig:isto_allerr}%
\end{figure*}
%
%
\section{Discussion and conclusions}
\label{concl}
We analysed the contribution of systematic uncertainties in the
diffuse Galactic emission surveys at 408, 1420 and 2326~MHz, cleaning the
striping effect arising in combination with the scanning strategy. 

We used the low frequency destriped maps to obtain the spectral index and 
normalization factor distributions over the whole observed sky. 

We performed a minimization technique over the calibration and zero level errors
of the surveys to reduce the discontinuities in the
$\alpha$ and $\beta$ maps in correspondence to the borders of the 1420 and 
2326~MHz maps. The artificial lines of discontinuities are reduced but not 
eliminated, suggesting for a complex pattern of $\delta_{\rm 0l}$ 
and $g_{\rm cal}$.

We discussed the spectral index distribution recovered; by estimating the
contribution of point sources and diffuse free-free emission, we concluded that
the spectral index distribution is representative for the diffuse 
Galactic synchrotron emission between 408 and 2326~MHz.

The new results of the $WMAP$ experiment give a quite precise picture of the 
Galactic emissions between $\simeq 23$ and 94~GHz (\cite{MAP_fore}).
The theory predicts a break in 
the synchrotron spectrum of the Milky Way at 22 GHz (\cite{voelk}), observed in the 
$WMAP$ data. 
In the synchrotron spectral index map between 0.408 and 23~GHz produced by the 
$WMAP$ team, emission from the
Galactic plane is dominant (the North Galactic Spur is much less evident than at
radio frequencies), since, with increasing frequency, the flat components of 
the spectrum become more important than the steep components. 
For example, the fact that in the $WMAP$ data the North Galactic Spur is much 
less evident than at radio frequencies shows clearly that any template of the 
Galactic emission is frequency dependent. This puts more interest in 
the comparison between the new data and those derived from radio 
surveys.

\begin{acknowledgements}

We warmly thank D.P. Finkbeiner for deep explanation of the destriping concept 
and Roger Hoyland for useful discussions.
Some of the results in this paper have been derived using the HEALPix
(G\`orski et al. 1999).
We wish to warmly thank the referee for useful and constructive comments.

\end{acknowledgements}

%
%

\end{document}